\input{aipcheck}

\documentclass[
    ,final            
  ]
  {aipproc}

\layoutstyle{6x9}

\begin{document}

\title{SECONDARY BARYON ASYMMETRY IN $\pi^{\pm} p$ COLLISIONS}


\author{G. H. Arakelyan}{
  address={Yerevan Physics Institute, Armenia\\
E-mail: argev@mail.yerphi.am}
}

\author{C. Merino,}{
  address={Departamento de F\'\i sica de Part\'\i culas, Facultade de 
F\'\i sica, and
Instituto Galego de Altas Enerx\'\i as (IGAE),
Universidade de  Santiago de Compostela, Galicia, Spain \\
E-mail: merino@fpaxp1.usc.es}
}

\author{Yu. M. Shabelski}{
  address={Petersburg Nuclear Physics Institute,
Gatchina, St.Petersburg, Russia\\
E-mail: shabelsk@thd.pnpi.spb.ru}
}

\begin{abstract}
The process of secondary baryon production in $\pi^{\pm} p$
collisions at high energies in the central and forward fragmentation regions 
is considered in the framework of the Quark-Gluon String Model.
The  contribution of the string-junction  mechanism to the 
baryon production is analysed. The results of numerical
calculations are in reasonable agreement with the experimental data on
the $\Lambda/\bar{\Lambda}$ and $p/\bar{p}$ asymmetries.
 


\end{abstract}

\maketitle


The Quark-Gluon String Model (QGSM) is based on the Dual Topological
Unitarization (DTU) and it describes quite reasonably  many features of 
high energy production processes, both
in hadron-nucleon and in hadron-nucleus collisions [1-6].
High energy interactions are considered as taking place via the exchange
of one or several Pomerons, and all elastic and inelastic processes
result from cutting through or between  those exchanged Pomerons 
\cite{AGK,TMPL44}. The
possibility of different  numbers of Pomerons  to be exchanged  introduces
absorptive corrections to the  cross-sections which are in
agreement with the experimental data on production of secondary hadrons. 

Hadrons are composite bound state configurations built up
from the quark and gluon fields. In the string 
models baryons are considered as configurations consisting of three strings
attached to three valence quarks and connected in one point that is
called string junction ($SJ$)~[9-11],
as it is shown in Fig.~1. Thus the $SJ$ mechanism 
has a nonperturbative origin in QCD. Such baryon structure is also supported by
lattice calculations \cite{Bornyanov}.
  
It is important to understand the role of the $SJ$ mechanism in
the dynamics of high energy hadronic interactions, in particular in
processes implying baryon number transfer~[13-15].
Significant results for hadron-hadron and
hadron-nucleus collisions were obtained in~[15-22]. 

In this note we analyse the contribution of the $SJ$ mechanism in the 
description of the existing data on 
spectra and on asymmetry of $\Lambda$ and $\bar{\Lambda}$ 
and on asymmetry of $p$ and $\bar p$ production. 
We use the same parametrisations of
diquark fragmentation functions to baryons and the same 
Regge trajectory intercepts as in~[15-19, 21, 22].

As it is thoroughly known the exchange of one or several Pomerons is one 
basic feature of high energy interactions in the frame of QGSM. Each Pomeron 
corresponds to a cylindrical  diagram, and thus when cutting a Pomeron two 
showers of secondaries are produced \cite{KTM,2r}. The inclusive spectrum of 
secondaries is determined by the convolution of diquark, valence, and sea 
quark distribution functions in the incident particle, $u(x,n)$, with the 
fragmentation functions of quarks and diquarks into secondary hadrons, 
$G^{h}(z)$. All these functions are determined by the corresponding Reggeon 
intercepts \cite{Kai}.

\vskip -1.cm
\begin{figure}[htb]
\centering
\includegraphics[width=.4\hsize]{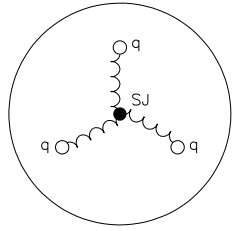}
\vspace{-2.5cm}
\caption{
Composite structure of a baryon in string model.}
\end{figure}

The formulas for inclusive spectra, i.e. the Feynman-$x$ distributions, of a 
secondary hadron $h$ in the QGSM together with the description of
high energy experimental data were presented in \cite{KTM,KPI,ACKS,AMSh}.
The analytical expressions for the complete set of the involved distribution 
and fragmentation functions are shown in ~\cite{AMSh}.

To produce a secondary
baryon in the process of diquark fragmentation three possibilities exist  
that are shown in Fig.~2.
The secondary baryon can consist of:
the initial $SJ$ together with two valence and one sea quarks (Fig.~2a),
the initial $SJ$ together with one valence
and two sea quarks (Fig.~2b), the initial $SJ$ together with three sea quarks 
(Fig.~2c). The fraction
of the incident baryon energy carried by the secondary baryon decreases from
case (a) to case (c), whereas the mean rapidity gap between the incident and
secondary baryon increases.

\begin{figure}[htb]
\centering
\includegraphics[width=.5\hsize]{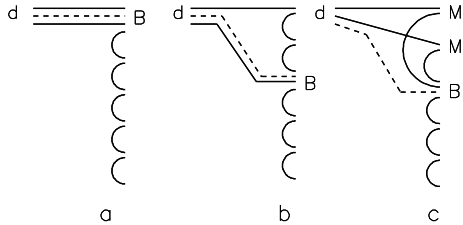}
\caption{
QGSM diagrams describing secondary baryon $B$ production by diquark 
$d$:(a) initial $SJ$ together with two valence quarks and
one sea quark, (b) initial $SJ$ together with one valence quark and two sea
quarks, and (c) initial $SJ$ together with three sea quarks.}
\centering
\end{figure}

The diagram of Fig. 2a describes the usual production of the leading baryon on 
diquark fragmentation.
The diagram of Fig. 2b has been used for the description of the baryon number
transfer \cite{KTM,22r}, and it also describes the fast meson
production by a diquark~\cite{Kai}.

\begin{figure}[htb]
\centering
\includegraphics[width=.45\hsize]{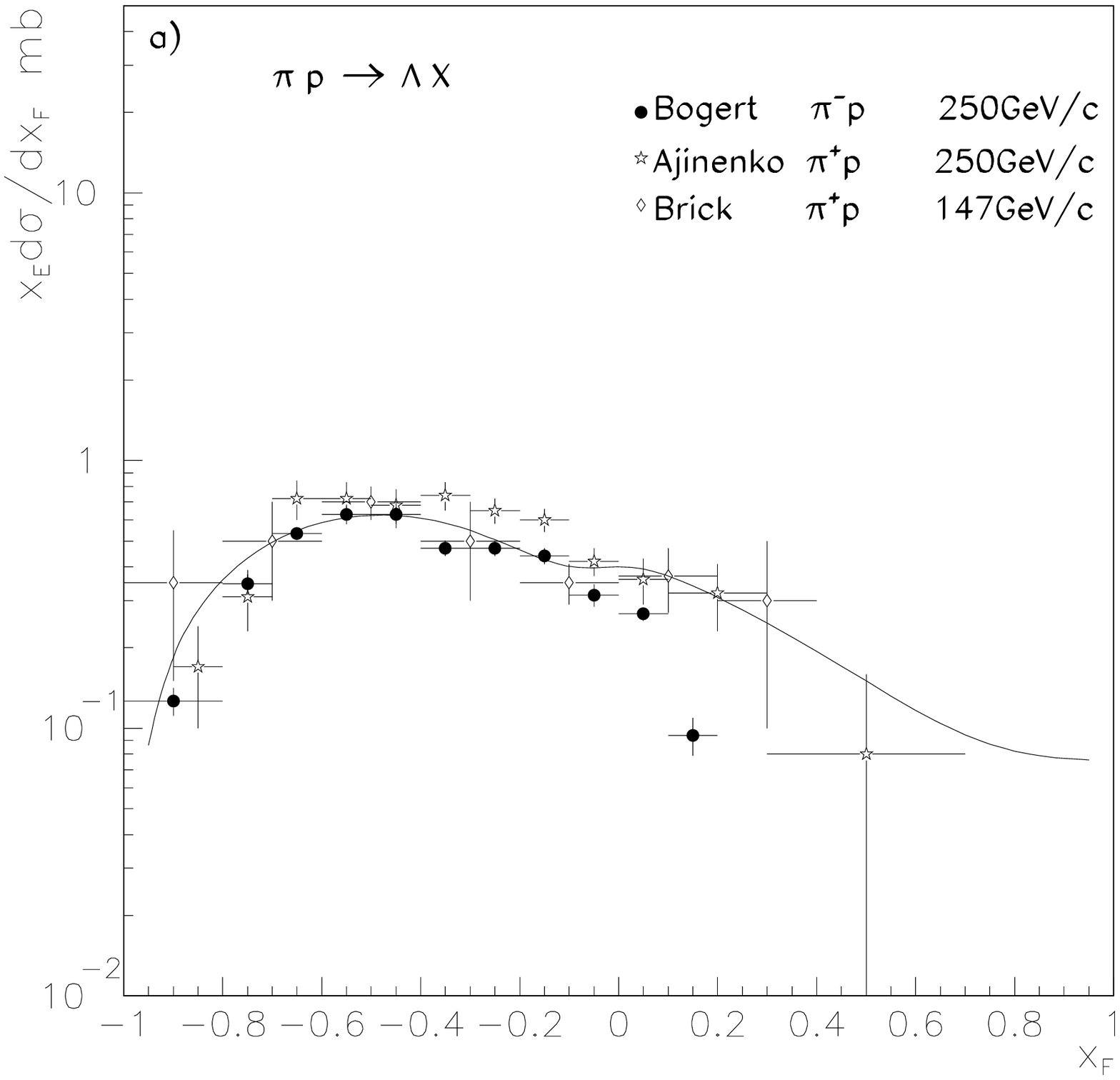}
\includegraphics[width=.45\hsize]{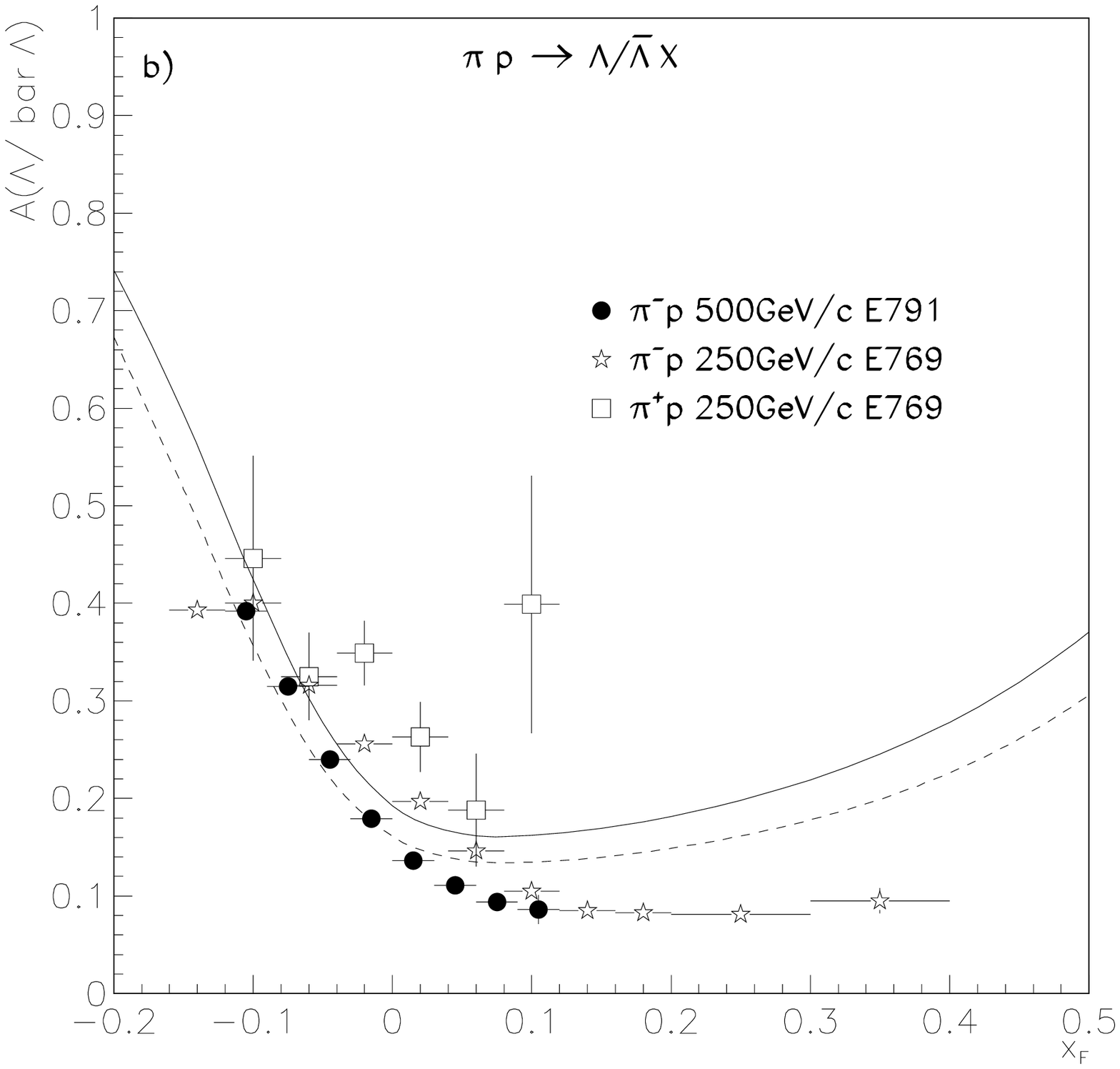}
\vspace{-0.5cm}
\caption{
The QGSM description of $\Lambda$ production and of the asymmetry
$A(\Lambda/\bar{\Lambda})$ in $\pi^{\pm} p$ collisions with $\alpha_{SJ}=0.9$: 
a) experimental data on the $\Lambda$ spectrum at 
$P_{lab}=147$~GeV/c ~\cite{Brick} 
and $P_{lab}=250$~GeV/c~\cite{Ajinenko,Bogert}; b) experimental data on  
asymmetry data at $P_{lab}=250$~GeV/c~\cite{Alves}
and $P_{lab}=500$~GeV/c ~\cite{ait1}. The solid curves correspond to the 
value $\delta=0.32$ and 
the dashed curve to $\delta=0.2$.}
\end{figure}

The contribution of
the graph in Fig. 2c to the diquark fragmentation function  is the most 
important for the
baryon number transfer to large rapidity distances. This contribution 
has been determined in~\cite{ACKS,ShBopp}, and it is proportional to a small  
coefficient $\varepsilon$ (the suppression factor of the process of
Fig. 2c compared to those in Figs. 2a and 2b).
The formulas for the diquark fragmentation functions corresponding to diagrams 
 on Fig.~2 are the following \cite {ACKS}:

\begin{equation}
\label{t13}
%
G_{uu}^p = G_{ud}^p = a_N z^{\beta}\Big[v_0\varepsilon (1-z)^2 +
v_q z^{2 - \beta} (1-z) + v_{qq}z^{2.5 - \beta}\Big] \; ,
\end{equation}
\begin{equation}
\label{t14}
 G_{ud}^{\Lambda} = a_N z^{\beta}
\Big[v_0\varepsilon (1-z)^2 + v_q z^{2 - \beta} (1-z) +
v_{qq}z^{2.5 - \beta}\Big](1-z)^{\Delta \alpha}\; , \;
G_{uu}^{\Lambda} =(1-z)G_{ud}^{\Lambda}. \;
\end{equation}
The terms proportional to $v_{0}$, $v_{q}$, and $v_{qq}$ correspond to the 
contributions of Fig. 2a,2b and 2c, respectively. 

The factor $z^{\beta}$ is actually $z^{1-\alpha_{SJ}}$, where $\alpha_{SJ}$ is
the intercept of the Reggeon trajectory which corresponds to the $SJ$
exchange. 
As for the factor $z^{\beta}\cdot z^{2 - \beta}$ in the second term,
it~corresponds to~$z^{2(\alpha_R - \alpha_B)}$~\cite{KTM}. For the third term
we have added an extra factor $z^{1/2}$~\cite{ACKS}.

The values of the parameters $v_{0}, v_{q}$, and $v_{qq}$  in 
Eqs.~(\ref{t13}) and ~(\ref{t14}) are obtained by simple quark combinatorics 
 \cite {ACKS,ANSH,CS}.

In the present calculation, we use the values of parameter
$\alpha_{SJ} = 0.9$ and of $\varepsilon=0.024$,
as it was done in~\cite{ShBopp,ShBnucl}.

The strange quark suppression factor $\delta$ in the model
is  usually taken for calculations in the interval $\delta \sim 0.2$-$0.3$.
As it was shown in \cite{ShBnucl}, the better agreement of QGSM with
experimental data on
strange baryon production on nucleus was obtained with $\delta =0.32$, instead
of the previous value $\delta =0.2$ \cite{ACKS}. In principle we cannot
exclude the possibility that the value of $\delta$ could be different for 
secondary baryons and mesons.

The $SJ$ mechanism does not affect the production of antibaryons, so 
the $\bar{\Lambda}$ spectrum doesn´t depend on the value 
of $\alpha_{SJ}$, and it has a very small dependence on the strange
 quark suppression factor $\delta$ (see \cite{AMSh,AMShk}).

In the figures the solid curves correspond to
calculations with $\alpha_{SJ}=0.9$, $\varepsilon=0.024$, and
$\delta=0.32$. Dashed curves are obtained with the same  $\alpha_{SJ}$ and 
$\varepsilon$, but with $\delta=0.2$.

The inclusive spectra of $\Lambda$ in $\pi^{\pm}p$ 
collisions at $P_{lab}=147$ GeV/c~\cite{Brick} and $P_{lab}=250$~GeV/c
~\cite{Ajinenko,Bogert} 
together with the model calculation are shown in Fig.~3a. Agreement is 
reasonably good, and the energy dependence of the model predictions is very 
weak at these energies. 

In Fig.~3b we show the data on the 
$A(\Lambda / \bar \Lambda)$ asymmetry 
\begin{equation}
\label{t15}
A(\Lambda/\bar{\Lambda}) = \frac{N_{\Lambda} - N_{\bar{\Lambda}}}{N_{\Lambda} + N_{\bar{\Lambda}}}
\end{equation}
in $\pi p$ interactions at $P_{lab}= 250$ GeV/c \cite{Alves} and 
$P_{lab}= 500$ GeV/c \cite{ait1}.

\begin{figure}[htb]
\centering
\includegraphics[width=.45\hsize]{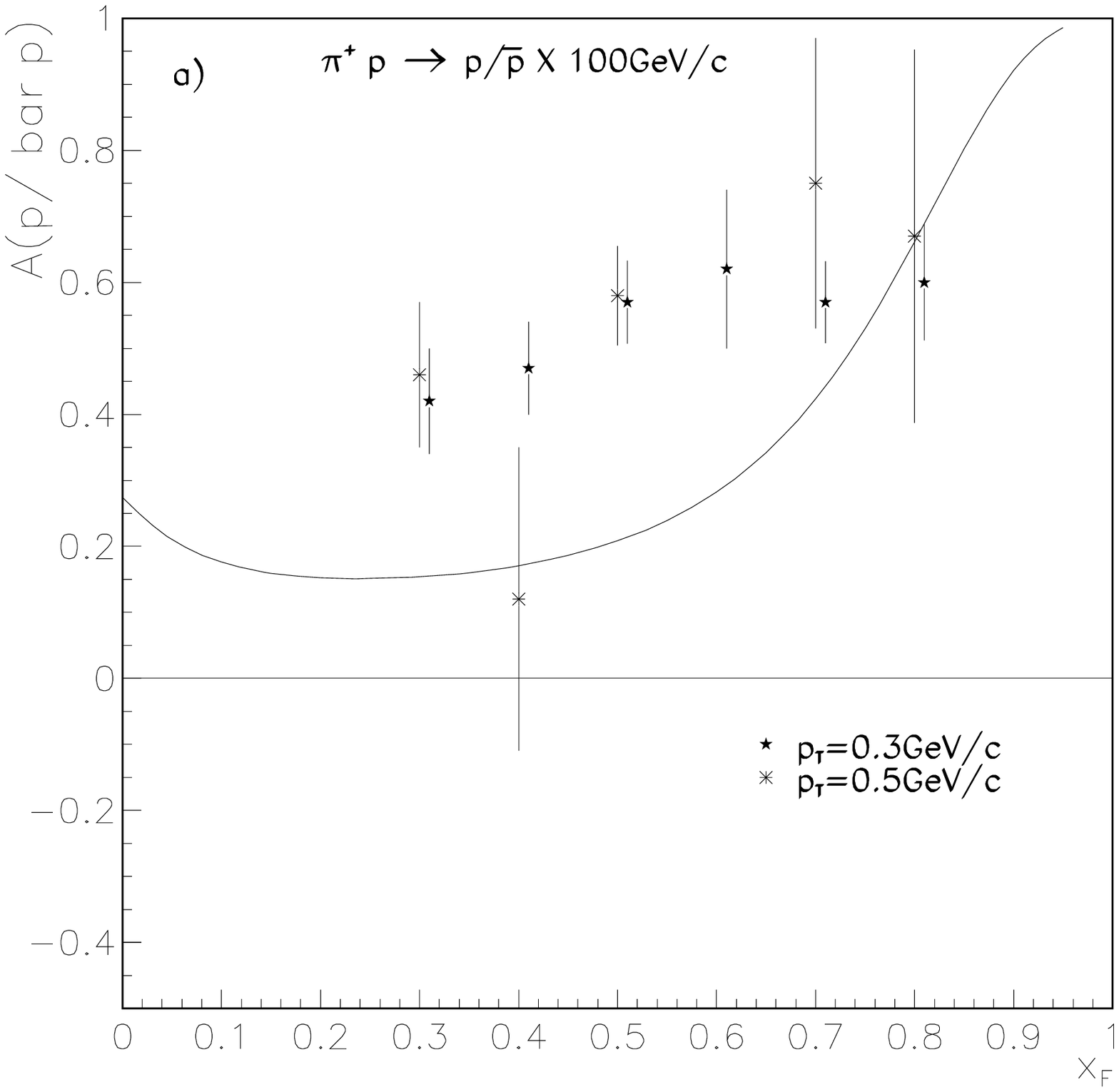}
\includegraphics[width=.45\hsize]{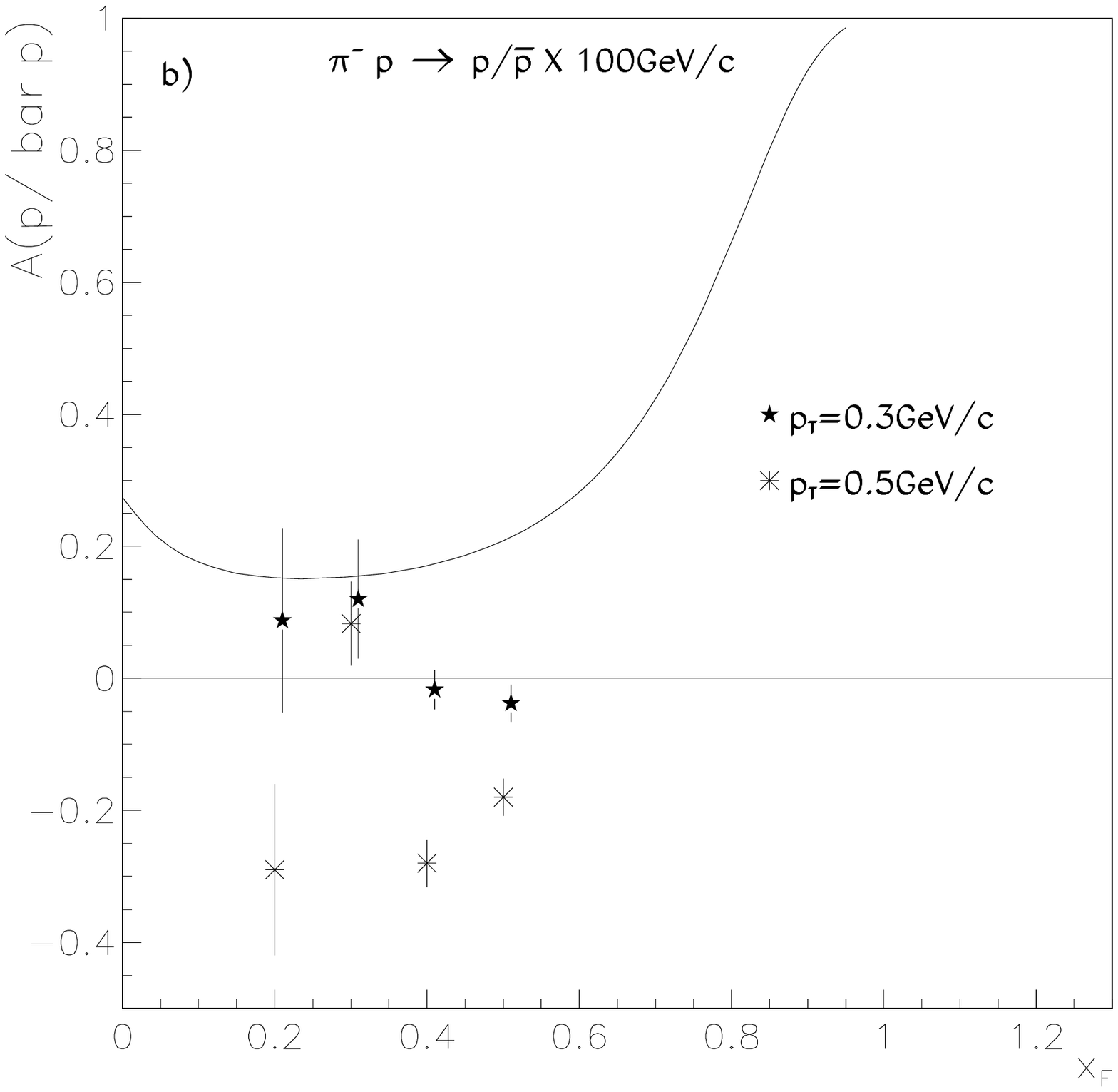}
\vspace{-0.5cm}
\caption{
The QGSM description of the asymmetry $A(p/\bar{p})$ in $\pi^+p$
and in $\pi^-p$ collisions
at $P_{lab}=100$~GeV/c. The solid curve corresponds to $SJ$ with
$\alpha_{SJ}=0.9$.}
\end{figure}

In the proton fragmentation region the  
values of $A(\Lambda/\bar{\Lambda})$ are close to unity, and that is natural
since a proton fragments into $\Lambda$ with significantly larger
probability than into $\bar{\Lambda}$. In the central and forward
fragmentation regions asymmetries are rather small. In principle, there is no
reason for a difference in the asymmetry $A(\Lambda/\bar{\Lambda})$ in 
$\pi^+$ 
and in $\pi^-$ collisions, since through isotopical reflection the reaction 
$\pi^+ p \to \Lambda$ becomes $\pi^- n \to \Lambda$, and vice versa, so the 
spectra of 
$\Lambda$ in $\pi^+ p$ and in $\pi^- p$ collisions should be similar. The same
can be said about $\bar{\Lambda}$ production. Actually in the QGSM frame the 
asymmetry $A(\Lambda/\bar{\Lambda})$  in $\pi^+ p$ and in $\pi^- p$
collisions is exactly the same. However, some difference appears in the 
experimental values of $A(\Lambda/\bar{\Lambda})$,
obtained in the same experiment E769~\cite{Alves}.
 
The inclusive spectra of secondary protons produced in $\pi p$ collisions at
 $P_{lab}=100$ GeV/c \cite{Brenner}, together with the
corresponding theoretical
curves, are presented in Fig.~4a ($\pi^+$ beam) and Fig.~4b ($\pi^-$ beam). 
As it was already noted, the  experimental data \cite{Brenner} contain only 
inclusive
spectra measured at fixed $p_T$. We calculate the corresponding asymmetry 
data where possible. As one can see, the comparison with the data is  
not good, in particular for $\pi^-$ beam. Here again the QGSM predicts rather
small difference in $A(p/\bar p)$ for $\pi^+ p$ and $\pi^- p$ collisionss, 
whereas the experimental behavior of the asymmetry for both beams is very 
different. The experimental data at $P_{lab}=175$ GeV/c \cite{Brenner} 
practically coincide with 
those in Fig.4 at $P_{lab}=100$ GeV/c. Possibly the disagreement is in part 
connected with the fact that in the case of 
$p/\bar p$ asymmetry only data at fixed $p_T$ are available.

In conclusion, the experimental data on high-energy $\Lambda$ production 
support the possibility of baryon charge transfer over large
rapidity distances. The $\bar{\Lambda}/\Lambda$ asymmetry is provided by
$SJ$ diffusion through baryon charge transfer.

To get a good understanding of the dynamics of the baryon charge transfer
over large rapidity distances new experimental data in meson and baryon
collisions with nucleons and nuclear targets are needed.


\begin{theacknowledgments}
The authors are thankful to A. B.~Kaidalov and C.~Pajares for useful
discussions.
 
This work was supported by Ministerio de Educaci\'on by Ciencia of Spain 
under 
project FPA2005-01963 and by Xunta de Galicia (Conseller\'\i a de Educaci\'on).
The work of Yu.M.Sh. were also supported in part by grant RSGSS-1124.2003.2.
G.H.A and Yu.M.Sh. thank the members of the Department of Particle Physics 
and of the Instituto Galego de Altas Enerx\'\i as~(IGAE), University of 
Santiago de Compostela, Galicia, Spain, for their kind hospitality.
\end{theacknowledgments}



\bibliographystyle{aipprocl} 

\bibliography{sample}



\end{document}